\documentclass[prl,letterpaper,onecolumn,showpacs,superscriptaddress,floatfix]{revtex4}

\usepackage{graphicx,psfrag,amsmath,amssymb,amsfonts,bbm,latexsym,color,dcolumn,bm}

\begin{document}

\title{Macroscopic quantum vacuum and microscopic gravitation}

\author{Roberto Onofrio}

\affiliation{Dipartimento di Fisica ``Galileo Galilei'', Universit\`a di Padova, Padova 35131, Italy}
\affiliation{Department of Physics and Astronomy, Dartmouth College, Hanover, NH
03766, USA}
\affiliation{ITAMP, Harvard-Smithsonian Center for Astrophysics, Cambridge, MA
02138, USA}

\begin{abstract}
Macroscopic quantum vacuum and modern theories of gravitation share the 
strong interplay between geometry and physical phenomena. 
We review selected issues related to the accuracy of the 
measurement of Casimir forces with particular emphasis 
on the implications for the search of non-Newtonian 
gravitational forces in the micrometer range.
We then discuss the interplay of the Higgs particle with 
gravitation, arguing that spectroscopic shifts in atomic 
transitions due to the modifications of the vacuum expectation 
value of the Higgs field in regions with strong curvature of 
space-time may be of conceptual and observational relevance. 
\end{abstract}


\maketitle

\section{Introduction}

Casimir forces have been extensively studied in the last 
decade in a variety of experimental configurations \cite{Mohideen,Lamoreaux}. 
At the level of demonstrating the existence of this 
peculiar manifestation of quantum vacuum at the macroscopic 
level, the situation seems rather settled. 
A demonstration of an effect which is expected since it is 
based on {\em the} prevailing theory, {\it i.e.} renormalized 
quantum electrodynamics, without predictions originating from 
alternative approaches \cite{Hagen}, seems rather uneventful and does not 
necessarily require an in-depth analysis of the data. 
Debate, however, is still going on to assess the precision and 
the accuracy of the measurements, since this has an impact on the  
potential discovery of new forces - predicted by a {\em variety} 
of models - with coupling comparable or larger than gravitation, and  
characterized by an interaction range at or below the micrometer scale
\cite{Giudice,Fujii,Fischbach}. 
In this framework, some previously unidentified systematic effects 
have been recently evidenced. If not properly taken into account 
in the data analysis and the experiment-theory comparison, neglecting
these effects may translate into significant systematic errors. 
In this contribution we briefly review the status of the experimental 
and theoretical knowledge in regard to such systematic effects, and the 
impact of the latter on the current limits to non-Newtonian gravitational 
forces. A complementary tool to extend our knowledge of quantum effects in 
the macroworld is available by considering peculiar quantum effects 
in strong external fields. In this framework we discuss the possibility 
to study observable effects due to the influence of curved space-time on 
the vacuum expectation value of the Higgs field. 
It turns out that the mass of the electron and, to a smaller extent,  
of baryons, should be changed in the presence of a strong gravitational
field. This in turn gives rise to spectroscopic shifts in atomic 
spectroscopy of a novel nature with respect to the usual Doppler, 
gravitational, or cosmological shifts, also raising conceptual 
puzzles about the assignment of masses via the Higgs mechanism.   

\section{Casimir force experiments and non-Newtonian limits in the
  micrometer range}

All experiments performed on Casimir forces between conducting bodies since 
the first attempts by Spaarnay \cite{Sparnaay} have faced the difficulty 
of dealing with a voltage present even when the two surfaces are shorted. 
This potential, called contact or residual potential, is thought to arise 
from the different Volta potential of the conductors used for the electrical 
connections, with typical values in the $1-100$ mV range. 
Apart from being superimposed to the expected Casimir force signal, its presence 
precludes the possibility to reach small gaps as it will cause, without taking 
proper precautions, earlier contact between the two surfaces. 
This is particularly relevant for experiments, such as the one 
using AFM cantilevers, involving resonators with low stiffness. 
It is then customary to compensate the contact potential using an external 
bias voltage kept constant in the entire range of explored distances. 
However, while performing electrostatic calibrations of an apparatus 
aimed at measuring the Casimir force in the cylinder-plane configuration, 
we have found evidence for a distance-dependent contact potential. 
Intrigued by this finding, we have studied the simpler sphere-plane 
configuration, since the latter is free from possible issues related 
to the parallelism and border effects potentially important in the 
cylinder-plane case. Even in the sphere-plane configuration, we have 
found a dependence of the contact potential upon distance, as well as  
an anomalous scaling of the electrostatic force with distance \cite{Kim1}. 
While this second anomaly does not seem of universal character \cite{Iannuzzi}, 
the dependence of the contact potential on distance has been confirmed 
by both reanalyzing former experiments (as discussed in \cite{JPCS}), and 
performing new experiments \cite{Iannuzzi,Kim2}. The presence of a 
distance-dependent contact potential creates an electrostatic force 
unevenly compensated at the various explored distances, unless on-line 
\cite{Iannuzzi} or off-line \cite{JPCS} compensation techniques are adopted.

\begin{table*}[b]
\begin{center}
\begin{tabular}{|l|c|c|c|}
\hline
 & Sphere-Plane & Cylinder-Plane  & Parallel Planes \\
\hline \hline
Casimir & $(\pi^3/360) \hbar c (R/d^3)$ & 
$(\pi^3/384\sqrt{2})\hbar c(La^{1/2}/d^{7/2})$ & 
$(\pi^2/240) \hbar c (S/d^4)$
\\ \hline
Coulomb & $\pi \epsilon_0 (R/d)V^2$ & 
$(\pi \epsilon_0/2\sqrt{2}) (La^{1/2}/d^{3/2})V^2$ 
& $(\epsilon_0/2) (S/d^2) V^2$ \\ \hline 
$V_{\mathrm{Cas}}^{\mathrm eq}$(d) & 
$(\pi^2/360)^{1/2} (\hbar c/\epsilon_0)^{1/2} 1/d$& 
$(\pi^2/120)^{1/2} (\hbar c/\epsilon_0)^{1/2} 1/d$   & 
$(\pi^2/192)^{1/2} (\hbar c/\epsilon_0)^{1/2} 1/d$   \\ 
\hline
$V_{\mathrm{Cas}}^{\mathrm{eq}}$ (1 $\mu$m) & 9.85 mV & 13.5 mV & 17.1 mV \\
\hline
\end{tabular}

\caption{Summary of relevant formulas for the ideal Casimir force and the 
Coulomb force in the cases of the sphere-plane, cylinder-plane, and 
parallel plane geometries. In the first row, the Casimir force is
expressed by regrouping the various numerical factors and variables 
in a common fashion, with numerical values first, then the product 
$\hbar c$ expected in any Casimir force formula, then the
geometrical dependence. The second row represents, in a similar 
arrangement, the Coulomb force exterted between the various surfaces.  
In the third row the equivalent Casimir voltage, {\it i.e.} the 
bias voltage required to simulate the Casimir force at a given 
distance $d$, is reported. In the last row the concrete value 
of the equivalent Casimir voltage is reported in the case of 
a typical gap distance of 1 $\mu$m. Notice the universality of 
the equivalent Casimir voltage formulas, with just different
numerical factors giving rise to a difference of less than 
a factor two between the two extreme geometries.}
\end{center}
\end{table*}

The situation remains far from being settled, since there are 
also several experimental results interpreted as if there is no distance 
dependence of the contact potential, within the precision of the 
measurement \cite{JPAMohideen,DeccaComment}. 
This issue is of the outmost importance to assess the best upper 
limits on Yukawa-like forces using molecular and Casimir forces 
\cite{Kuzmin,MosteSok}, as well as claims of validation of one 
specific model to include the finite conductivity and finite 
temperature correction \cite{Deccannal}.
Indeed, both the upper limits on Yukawa forces and the validation 
of thermal corrections rely on the two experimental apparata in which 
no systematic dependence of the contact potential on distance is observed 
(see \cite{RMP} for details). 
The effect observed in the other experiments is of the order 
of 4-10 mV in a range of distances of few $\mu$m. 
This voltage should be compared with the equivalent voltage 
corresponding to the Casimir force at a given distance \cite{OnCa}. 
When this is done, a simple and universal formula, apart from a 
numerical factor, emerges for the equivalent voltage of the three 
geometries. As seen in Table I, the equivalent voltage of the 
Casimir force at 1 $\mu$m is in the 10-20 mV range depending 
on the specific geometry, and this implies that the expected 
signal at 1 $\mu$m for the Casimir force may be comparable, within 
one order of magnitude, to the difference between the signals coming 
from the contact potentials at the two extremes of the explored range 
of distances. 

The current setting is clearly unsatisfactory to the purpose of providing 
reliable and accurate limits to non-Newtonian forces, regardless of the 
extent of the systematic effect. Indeed, even without carrying out extensive 
assessments of the precision of the experiments, we can identify three possible scenarios.
If this previously unidentified systematic effect is larger than 
the quoted systematic error in the experiments used to give the 
best limits on Yukawa forces, then limits should be revised taking 
into account this systematic source of error. 
If it is instead smaller than the quoted systematic error in the 
same experiments, then one needs to understand whether these 
experiments, unable to evidence such an effect unlike all 
others, are actually suitable to provide the best limits 
to non-Newtonian gravitation. 
Finally, there is the possibility that specific environmental 
factors varying from experiment to experiment, such as the 
geometrical quality and chemical contamination of the surfaces, the 
radius of curvature of the sphere, the vacuum level, specific electrostatic 
setting around the apparatus, and temperature control for instance, may 
explain the different observations. Moreover, the same intrinsic 
geometry of the Casimir cavity seems to be relevant, as there 
is no evidence so far of the dependence of the contact potential 
on distance in the case of the parallel plate configuration 
\cite{CQG01,PRL02,JPCS09}, as we will report in a future publication. 
Evidently more experimental and phenomenological work in this 
direction will be required to provide a consistent framework.

A second issue only addressed quite recently is the validity of 
the proximity force approximation (PFA) \cite{Derjaguin} for 
Yukawa or, in general, volumetric forces. Since the PFA was 
conceived to deal with proximity forces, {\it i.e.} forces acting 
among entities in proximity of each other, it is not {\it a priori} 
understood what is their range of validity, and the level of accuracy, 
whenever PFA is applied to forces acting among entities in the bulk. 
A simplified form of PFA based on a virtual mapping between the actual sphere-plane 
configuration used in the experiments and an effective parallel plane 
configuration, used for instance in \cite{Deccannal}, has been recently 
shown to differ from the general expression for the PFA to be used in 
the case of volumetric forces \cite{DeccaPFA}. 
This latter form has been in turn shown to coincide with the exact 
force between two bodies, provided one is an indefinite plane \cite{PFAYuk}. 
In the same paper, the sensitivity of the simplified PFA approximation to 
an unphysical parameter used in the PFA mapping between the
sphere-plane and the parallel plane case, the thickness of the 
hypothetical slab corresponding to the sphere, denoted $D_2$ 
in \cite{DeccaPFA} and \cite{PFAYuk}, has been also studied. 
Taking into account the dependence on this parameter, limits derived 
through the simplified PFA become quickly unreliable as the Compton wavelength 
of the Yukawa force approaches the micrometer range, {\it i.e.} the upper 
range of distances in which precision Casimir force measurements have 
been performed. Then the validity of the PFA in the case of volumetric 
forces of Yukawa type is limited not only by the usual constraint of 
distance being much smaller than the radius of curvature of the sphere, 
$a \ll R$, but also by Yukawa range $\lambda \ll R, D_2$. 
This limits its conceptual validity to the case of nearly pointlike 
interactions and, above all, to a region of distances smaller than 
the one in which actual experiments are performed.
This issue is not of practical concern for current experiments due to 
the dominance of the precision (of order 1.0 $\%$ in the explored 
range of distances, see Fig. 10 in \cite{Deccannal}) with respect 
to the estimated approximation in using the PFA formula for the 
Casimir force rather than its exact expression (of order 0.1 $\%$).
As the experimental precision will improve, for instance by developing 
low-temperature experiments in the submicrometer range, in analogy to 
the experiment described in \cite{Mullin} with attoNewton sensitivity, 
and if the sphere-plane PFA formula for the Casimir force will be replaced 
by its exact expression, the use of the exact Yukawa force will become 
increasingly crucial. The exact expression for the Yukawa force in the 
sphere-plane geometry, already available since a decade \cite{Bordag1998}, 
has been recently used to improve the limits in the micrometer range 
using a torsional balance \cite{Masuda1,Masuda2}.

\section{Higgs shifts in astrophysical environments}

One of the most important predictions of the standard model of particle 
physics is the existence of the only fundamental scalar particle held 
responsible for the spontaneous symmetry breaking of the electroweak 
sector, providing mass to the intermediate vector bosons $W^{\pm}$ 
and $Z^0$ and to all fundamental fermionic matter fields. 
The Higgs particle is a critical milestone of the standard model, 
and its discovery and detailed study is the primary focus of research 
undergoing at high energy accelerators such as Fermilab and, 
in the close future, at the Large Hadron Collider at CERN. 
The Higgs particle, if giving mass to all the constituents of matter 
and provided that it satisfies the equivalence principle, should 
also play a crucial role in gravitational phenomena. 
In particular, the Higgs vacuum expectation value in a region of 
strongly curved space-time should differ from the one in flat space-time. 
This should give rise to different values for the mass of particles 
such as electrons and protons, and then to shifts of energy levels 
of their bound states of spectroscopic relevance. 
New wavelength shifts in the emission or absorption spectra are then 
predicted, which could be detected by proper subtraction of the usual 
Doppler, gravitational, and cosmological shifts.    

Quantum field theory in curved space-time has been studied for 
decades both for non-interacting and interacting fields \cite{Birrell}.
The Lagrangian density for an interacting scalar field with parameters 
$\mu$ and $\lambda$ in a generic curved space-time is written as:
\begin{equation}
L= \frac{1}{2} g^{\mu \nu} 
\partial_\mu \phi \partial_\nu \phi - \frac{1}{2} (\mu^2 +\xi R) \phi^2 
-\frac{\lambda}{4}\phi^4.
\end{equation}
In the presence of a curved space-time there is an added term to 
the Lagrangian density with respect to the case of a flat space-time, 
where $\xi$ measures the coupling between the Higgs field $\phi$ and 
the curvature scalar $R$. In the {\em minimal} coupling scenario, 
we should have $\xi=0$, which however is unnatural if we believe 
that the standard model at some energy will merge with gravitation, 
since it precludes any possible crosstalk between the two sectors.
Moreover, it does not manifest proper renormalization group behavior, as 
outlined in \cite{Birrell} and, in a composite scalar model, in 
\cite{Hill}. Alternatively, we consider as a working 
and/or reasonable hypothesis the other possibility of a {\em conformal} 
coupling, $\xi=1/6$. In the spontaneously broken phase, the Higgs field 
develops a vacuum expectation value obtained by minimizing the effective 
potential. In the flat space-time this yields a value 
$v_0=\sqrt{-\mu^2/\lambda}$, and the masses of the elementary 
particles are all directly proportional to $v$ via the Yukawa 
coefficients $y_i$ of the fermion-Higgs Lagrangian density term, 
$m_i=y_i v_0/\sqrt{2}$. In the presence of curved space-time, 
the effective coefficient of the Higgs field $\mu^2 \mapsto \mu^2+\xi R$ 
and the vacuum expectation value of the Higgs field will become 
space-time dependent through the curvature scalar as:
\begin{equation}
v=\sqrt{-\frac{\mu^2+\xi R}{\lambda}} \simeq v_0 \left(1+\frac{\xi R}{2 \mu^2}\right),
\end{equation}
where the last expression holds for a weak curvature. 
Notice that the vacuum expectation value is increased by the presence
of a curved space-time corresponding to $R>0$ in the minimal coupling scenario. 
In the case of elementary particles such as the electrons, provided
that the Yukawa couplings are constants yet to be determined -  
as commonly believed,  from algebraic/group theoretic arguments 
of an underlying fundamental theory incorporating the standard model - 
the mass $m_e$ will be simply changed proportionally to the Higgs 
vacuum expectation value, so that in the case of the electron 
$\delta m_e=y_e (v-v_0)/\sqrt{2}\simeq y_e \xi R v_0/(2^{3/2} \mu^2)=m_e \xi R/(2 \mu^2)$. 

The situation for composite particles such as protons and neutrons is more 
involved. We assume that their masses are made of a flavor-dependent 
contribution proportional to the masses of the three valence quarks
determined by the Higgs coupling, and a color-symmetrical term only 
dependent on the quark-quark and quark-gluon interaction, {\it i.e.} 
proportional to $\Lambda_{QCD} \simeq 300$ MeV. The latter dominates 
for lighter, relativistic quarks such as the up and down 
quarks constituting the valence component of protons and
neutrons. Then, due to the universality of the QCD coupling constant 
for different flavors and for all gluons exchange, the QCD-related mass 
term will be the same for protons and neutrons, and therefore we will have:
\begin{equation}
m_p = (2 y_u+y_d) v/\sqrt{2}+m_{QCD}, ~~~ m_n = (y_u+2y_d) v/\sqrt{2}+m_{QCD}.
\end{equation}
For a generic atom of atomic number $Z$ and atomic mass $A$ we obtain:
\begin{equation}
M(A,Z) =\frac{1}{\sqrt{2}}[y_u (Z+A)+y_d(2A-Z)]v+Am_{QCD},
\end{equation}
where we have neglected, to first approximation, the electron mass, the 
electron-nucleus binding energy and the nucleon binding energy. 
The purely QCD-dependent mass term should be independent on the 
curvature of space-time, since otherwise the gluon could acquire 
a mass giving rise to the explicit breaking of the color symmetry. 
This is analogous to the case of the other unbroken symmetry of 
the standard model, $U(1)_{em}$ leading to the electromagnetic 
charge being conserved even in a generic curved space-time. 

The possibility to detect Higgs shifts in atomic and molecular
spectroscopy relies on the fact that electronic transitions 
depend primarily on the mass of the electron, whereas molecular 
transitions due to vibrational or rotational degrees of freedom 
depend upon the mass of the nuclei. While the electron mass is 
directly proportional to the Yukawa coupling, the mass of the 
nuclei is mainly due to the mass of its proton and neutron 
constituents, which in turn depends mainly on the color binding 
energy. We therefore expect that molecular transitions will not be
affected by the Higgs shifts at leading order, unlike electronic 
transitions, not even in the most sensitive case of pyramidal molecules
such as ammonia, for which tunneling provides enhanced sensitivity to 
changes in masses of the nuclei for the inversion lines. 
It is difficult to detect electronic, vibrational, and rotational 
transitions in the same region of space from the same species 
for a gas at thermal equilibrium, due to the very different energy scale 
difference required to produce these excitations. A comparative
analysis of wavelength shifts from different species seems then 
necessary. This also enables to disentangle the putative Higgs 
shift contribution from the Doppler shift and the purely gravitational 
shift. The Doppler shift should be the same for molecules belonging to 
the same comoving cloud, while the wavelength shift expected from 
general relativity will act universally on all particles. Thus, unlike 
the Higgs shift, the latter will not discriminate between fundamental particles 
and interactions binding energies, {\it i.e.}, between the electron 
mass and the main contribution to the proton mass due to QCD. 
A promising object to look for possible Higgs shifts is the Galactic 
center, where a compact object with an estimated mass of 
$M \simeq 2.6 \times 10^6$ solar masses, Sagittarius A$^*$ 
\cite{Morris,Eckart,Genzel,Ghez, Reid}, is present and supposed 
to be a black hole with a Schwarzschild radius of $R_s=2 G M/c^2 \simeq 4 \times 10^9$ m. 
Unfortunately, the curvature scalar is $R=0$ for a Schwarzschild black hole.
Under the hypothesis that the Higgs field is coupled to another 
scalar invariant, for instance the Kretschmann invariant 
$K_1=R_{\mu\nu\rho\sigma}R^{\mu\nu\rho\sigma}$, then we have 
$\mu^2 \mapsto \mu^2+\xi' {K_1}^{1/2}$. 
For the Schwarzschild metric $K_1=12 R_s^2/r^6$, where $r$ is 
the distance from the center of the mass, and therefore we obtain 
$\mu^2 \mapsto \mu^2 (1+ 2 \sqrt{3} \xi' R_s \lambda_{\mu}^2/r^3)$, 
where $\lambda_{\mu}=\hbar/(\mu c)$ is the Compton wavelength 
associated to the Higgs mass parameter $\mu$. Assuming a Higgs mass 
of 160 GeV, with $v=247$ GeV, we obtain a Compton wavelength 
$\lambda_{\mu} \simeq 2 \times 10^{-18}$ m. 
Even if we consider the innermost stable orbit around 
a solar mass black hole, at $r=3 R_s$, we obtain 
$K_1^{1/2}={(4/243)}^{1/2} R_s^{-2}=1.5 \times 10^{-8}$ m${}^{-2}$.
Therefore the product of the squared Compton wavelength and the Kretschmann 
term is $\lambda_{\mu}^2 K_1^{1/2} \simeq 6 \times 10^{-44}$, leading 
to Higgs shifts very far from observational reach. The main issue 
here is the mismatch in the lengthscales $\lambda_{\mu}$ and 
$K_1^{-1/4}$ which could be reduced if mini black holes are considered.
It should be also noticed that the lack of detailed knowledge of
compact objects cannot rule out their possible description in terms 
of naked singularities, providing effective curvatures better matched 
to $\lambda_{\mu}$, although the very existence of naked singularities 
is a hotly debated theme in general relativity \cite{Joshi}. 
Upper bounds on the existence of Higgs shifts from dedicated surveys 
can be anyway useful to constraint proposals in which the Higgs boson 
is supposed to be responsible for inflation via anomalously large coupling  
to the metric, as proposed in \cite{Bezrukov1,Bezrukov2}.

As benchmarks from the observational viewpoint, with a 1 pc resolution 
survey it is now possible to obtain spectra of atoms or molecules at 
a distance of $r \simeq 2 \times 10^{16}$ m from the Galactic center, 
and recent surveys of ammonia \cite{Henkel,Wilson} have a spectral 
sensitivity corresponding to a Doppler shift of about 2-3 km/s, 
{\it i.e.} $\delta \nu/\nu \simeq 10^{-5}$. 
In the case of atomic hydrogen spectroscopy, the spectral lines depend 
directly on the reduced mass $\mu_{ep}=m_e m_p/(m_e+m_p)$ and ultimately, due to the 
large mass ratio $m_p/m_e$, on the electron mass. Observation of
atomic lines from the Galactic center is difficult due to the strong 
absorption at optical wavelengths, and therefore one should focus on 
the high-precision monitoring of the 21-cm neutral hydrogen line, still
dependent on the electron to proton mass ratio.   
This implies measuring the 21-cm neutral hydrogen line from interstellar 
clouds near the Galactic Center, or from stars with highly eccentric orbits. 
In the latter case, clear signatures might be available by looking at 
the temporal variability of the 21 cm hyperfine line along the star orbit. 

Finally, we want to point out that in the standard model the 
masses of fundamental particles have a different treatment with 
respect to the mass coming from interaction among themselves. 
If the equivalence principle holds, the gravitational mass 
of the electrons constituting a test body will change if the 
Higgs field is coupled to curvature, while the nucleons will 
continue to keep, to leading order, the usual gravitational charge.
This is in striking contrast to the case of general relativity where 
all sources of energy contribute without any distinctive 
feature to the emergence of space-time, originating at least 
an unappealing contrast in the way masses are considered 
in Higgs physics and gravitation. Stability issues of back-reaction 
on the metric may also arise depending on the relative signs of 
$\xi$ and $R$, potentially originating catastrophic growths of 
the local curvature and of the mass of the test particles.

\section{Conclusions}

We have briefly discussed two critical issues in the interplay between 
macroscopic quantum vacuum and microscopic gravitation, namely the need 
for a more careful analysis of systematic effects in Casimir force 
measurements to improve our knowledge of possible non-Newtonian 
forces of gravitational origin, and the influence of strong space-time 
curvature on the generation of mass generation via the Higgs field.
By smearing out the two traditional regions of applicability of 
quantum physics and gravitation, the microworld and the macroworld
respectively, one could gain important insights on their mutual compatibility, 
as already attempted in \cite{Karim1,Karim2,Calloni,Sorge,Milton1,Milton2,Milton3,Milton4} 
for the relationship between vacuum energy and curved space-time, 
and in \cite{Viola,Onofrio} concerning the validity of the equivalence 
principle for macroscopically distinguishable quantum states.
All this should be considered part of a program aimed at narrowing 
the gap between the physics of quantum vacuum and the standard model of 
elementary particle physics and its proposed extensions, as recently 
outlined in \cite{Gies}.
 
\section*{Acknowledgments}
I would like to thank the QFEXT09 organizers for the kind invitation to such 
a stimulating and successful meeting. I also acknowledge partial support 
from the Julian Schwinger Foundation through grant JSF 08070000 on Astrophysics 
of Quantum Vacuum.

\end{document}